\RequirePackage{fix-cm}
\documentclass[]{relaxed_system_lab}

\ifPDFTeX
\usepackage[utf8]{inputenc}
\fi
\usepackage[T1]{fontenc}
\usepackage{charter}
\usepackage{xspace,flafter,placeins}
\usepackage{needspace}
\usepackage{amsmath,amssymb}
\usepackage{graphicx,booktabs,tabularx,array}
\usepackage{xcolor,microtype,xurl}
\usepackage{enumitem}

\newcommand{\sys}{$e$\textsc{PACT}\xspace}

\newif\ifdraftnotes
\draftnotestrue
\newif\ifappendixdetails
\appendixdetailsfalse
\newcommand{\lead}[1]{\noindent\textbf{#1.}}
\newcommand{\pos}[1]{\left[#1\right]_{+}}

\newcolumntype{Y}{>{\raggedright\arraybackslash}X}
\DeclareMathOperator*{\argmin}{arg\,min}

\appendixdetailstrue 

\hypersetup{
  pdftitle={ePACT: Energy-Performance-Aware Commitment Tracking for LLM Serving},
  pdfauthor={You Peng, Youhe Jiang, Chen Wang, Binhang Yuan}
}
\title{\sys: Energy-Performance-Aware Commitment Tracking for LLM Serving}
\author{You Peng$^1$$^*$, Youhe Jiang$^1$$^*$, Chen Wang$^2$, Binhang Yuan$^1$}
\affiliation{$^1$HKUST, $^2$Tsinghua University}
\contribution{$^*$Equal contribution}

\abstract{Reducing LLM serving energy does not by itself guarantee lower deployment cost when electricity procurement exposes operators to unfavorable deviations from preset commitments. We study hourly commitments with positive, potentially asymmetric costs for overuse and underuse, and formulate \underline{e}nergy-\underline{P}erformance-\underline{A}ware \underline{C}ommitment \underline{T}racking: minimize deviation costs subject to request-level service requirements. We implement \sys, a two-level controller that adjusts serving capacity and GPU clocks as requests arrive. A global planner updates interval energy targets from measured consumption and the remaining hourly commitment. A local decision maker predicts candidate configurations' energy and completion times, checks predicted deadline misses, and selects among admitted configurations by asymmetric target-deviation cost, with a service-first fallback. Coarse-to-fine action search runs asynchronously with serving. We evaluate \sys through single-hour comparisons, controller ablations, and full-day trace simulations for H20 and H200 GPU pools. In the 24-hour simulations, \sys reduces the asymmetric deviation cost by $73.8\%$ and $75.7\%$ relative to vLLM while retaining near-vLLM SLO attainment. Mean absolute hourly deviations are $2.16\%$ and $2.31\%$, respectively.
}

\begin{document}
\maketitle

\section{Introduction}
\label{sec:intro}

LLM serving increasingly contributes to large, variable data center electricity demand \citep{lbnl2024}. The International Energy Agency estimates that data centers consumed $485$\,TWh in 2025 and projects approximately $950$\,TWh in 2030 \citep{iea2026}. Operators are also incorporating computing flexibility into energy procurement: Google reports integrating $1$\,GW of data center demand response into long-term utility contracts \citep{google2026}. Reducing physical energy consumption alone does not guarantee an economic benefit, because serving decisions must also account for electricity commitments and service obligations. This paper asks: \textit{How can an online LLM serving system minimize the costs of overusing or underusing preset hourly energy commitments while maintaining request-level Service-Level Objectives (SLOs) under uncertain demand?}

Advance electricity procurement connects serving demand to decisions made before requests arrive. Buyers can establish financially binding energy positions before delivery, as in PJM's hourly day-ahead market \citep{pjm_energy,pjm_billing}; the expected demand helps the grid schedule generation and maintain real-time balance \citep{eia_grid}. In our setting, an operator supplies hourly energy commitments before online serving begins, using historical workloads to estimate demand and the energy needed to meet its service requirements. Once these volumes are fixed, changing a serving configuration changes actual consumption but does not revise the commitment. Consuming more than contracted can create additional settlement exposure, while consuming less can also incur charges under volume-tolerance provisions \citep{elexon_imbalance,wapa_imbalance,edf2015}. The signs, rates, and settlement periods are contract-specific. We focus on the regime with positive costs for both overuse and underuse, modeled at hourly granularity with potentially different rates for the two directions. In this regime, the economic consequence of a serving decision depends on its effect on the gap between realized and committed energy, not solely on the energy it saves.

This distinction changes which serving actions are desirable. Suppose consumption is falling behind the committed trajectory while requests are waiting. Increasing active capacity or GPU frequency can then serve useful work sooner and reduce underuse exposure, even though it consumes more energy. An energy-minimizing policy can miss this opportunity by favoring a smaller configuration. Conversely, when consumption is ahead of the committed trajectory, reducing capacity or frequency can limit overuse if sufficient deadline slack remains. The same workload therefore calls for different actions depending on the energy already consumed and the commitment still to be fulfilled. A fixed power cap cannot express this dependence on cumulative energy, and low daily consumption can hide costly deviations in individual hours. Economically effective serving in this regime requires \emph{commitment-aware control}: minimize asymmetric deviation costs while protecting request-level service quality. The commitment serves as a cost reference rather than a hard cap or an instruction to generate unnecessary work. Procurement and tariff selection remain outside the controller; its task is to use serving flexibility to pursue the supplied commitment despite uncertain arrivals and limited SLO slack.

Tracking these commitments online raises two challenges. \underline{Challenge 1:} \textit{Workload uncertainty.} The system must serve incoming requests while tracking the current hour's energy commitment without knowing future arrivals. Each serving decision changes both cumulative energy consumption and the backlog carried into subsequent time intervals, so later decisions must account for both earlier consumption and the remaining work. 
\underline{Challenge 2:} \textit{Service quality under two-sided energy costs.} Favoring low latency can consume energy too aggressively and overshoot the commitment, whereas aggressively reducing energy use can increase SLO violations and underuse costs. The useful direction of energy adjustment therefore depends on the remaining commitment as well as the workload; serving decisions must protect service quality while limiting both forms of deviation.

The performance and energy objectives of existing LLM serving systems do not directly address this two-sided tracking problem. Systems such as vLLM, Sarathi, DistServe, and Splitwise improve serving efficiency through memory management, batch scheduling, and phase separation \citep{vllm,sarathi,distserve,splitwise}, while energy-aware systems such as DynamoLLM and throttLL'eM reduce inference energy subject to performance constraints \citep{dynamollm,throttllem}. These objectives favor lower latency or lower consumption, rather than alignment with a supplied hourly commitment. Neither objective alone resolves the joint risk of SLO violations and deviations on either side of that commitment.

We introduce \sys, which divides each hourly settlement horizon into decision intervals and couples \emph{interval energy targets} with \emph{serving-configuration selection}. To tackle \underline{Challenge 1}, a global planner updates targets from measured consumption and the remaining hourly commitment. To tackle \underline{Challenge 2}, a local decision maker evaluates candidate configurations from the current serving state, checks predicted deadline misses, and ranks admitted candidates by asymmetric deviation cost relative to the interval target. Actual consumption closes the loop at the next boundary. Adaptive action search first samples the configuration space broadly, then refines promising neighborhoods while requests continue to execute.

In summary, we make the following contributions:
\begin{itemize}[leftmargin=*]
  \item We formulate online LLM serving under preset hourly energy commitments as a two-sided tracking problem: minimize asymmetric overuse and underuse costs subject to a prescribed deadline-miss tolerance, accounting for energy across active, draining, and standby replicas without knowing future arrivals.
  \item We design \sys, a two-level controller that adjusts interval targets using measured consumption and jointly selects replica counts and GPU clocks using snapshot-based SLO predictions. Coarse-to-fine search reuses promising configurations while serving continues asynchronously.
  \item We evaluate \sys through single-hour comparisons, controller ablations, and 24-hour trace simulations for H20 and H200 pools. In the full-day simulations, it reduces the asymmetric deviation cost by $73.8\%$ and $75.7\%$ relative to vLLM, with near-vLLM SLO attainment and mean absolute hourly deviations of $2.16\%$ and $2.31\%$, respectively. The ablations characterize the contributions of global feedback and local configuration adaptation.
\end{itemize}

\section{Preliminaries \& Related Work}
\label{sec:preliminaries}
\label{sec:related}

We review how serving mechanisms and SLOs constrain configuration choices, then distinguish commitment tracking from existing energy objectives.

\noindent\textbf{LLM serving under SLOs.}
An autoregressive request first processes its prompt in prefill and then generates output tokens through repeated decode steps. Iteration-level scheduling admits and retires requests as generation proceeds \citep{orca}, while vLLM's PagedAttention improves the management of their key-value caches \citep{vllm}. Sarathi-Serve uses chunked prefill to reduce interference with decoding \citep{sarathi}; DistServe separates the two phases to optimize goodput under time-to-first-token (TTFT) and time-per-output-token (TPOT) constraints \citep{distserve}. Serving quality is commonly expressed through latency SLOs, where a request is compliant when its latency remains within a prescribed threshold \citep{clockwork,llumnix}. To account for heterogeneous request lengths, prior inference-serving work also uses scaled SLOs, which set the latency threshold relative to a baseline or isolated execution latency through a multiplicative SLO scale \citep{alpaserve,sparsedysta,hexgen,hexgen2,adaserve}. We use request-level deadlines to express this service requirement while controlling serving capacity.

\nocite{jiang2025demystifying,jiang2025thunderserve,jiang2026oserve,jiang2026cascadia,tong2025parallax,peng2026hexgen,jianghexgen}

\noindent\textbf{Electricity energy commitment.}
A power cap constrains instantaneous demand, whereas an energy commitment concerns cumulative consumption over a settlement horizon. Volume-tolerance provisions can charge for deviations in either direction, with contract-specific periods and tolerances \citep{edf_academy,edf_flexible}. We model hourly commitments with positive two-sided deviation costs; under this assumption, lower energy consumption need not mean lower cost. Related carbon-aware systems shift flexible computation in time. Google's Carbon-Intelligent Compute Management \citep{carbonaware2023} uses Virtual Capacity Curves to shift flexible workloads, while the probabilistic carbon-aware scheduling framework \citep{hall2025} combines day-ahead planning with real-time tracking to reduce carbon cost and peak power. Our objective instead minimizes deviation costs for a supplied hourly commitment under request-level service requirements.

\noindent\textbf{Energy-aware LLM Inference service.}
DynamoLLM coordinates instance count, parallelism, and frequency to reduce energy under SLOs \citep{dynamollm}. throttLL'eM predicts serving behavior to scale instances and GPU frequency \citep{throttllem}, while GreenLLM applies SLO-aware frequency control to prefill and decode \citep{greenllm}. For prefill-decode disaggregated serving, VoltanaLLM combines frequency control with routing \citep{voltanallm}, and DualScale couples phase-aware placement with fine-grained frequency control \citep{dualscale}. These studies establish the value of adapting capacity and frequency. \sys uses these controls to reduce deviation from an interval target, including cases where the preferred SLO-admitted configuration consumes more energy.

\noindent\textbf{Online scheduling and simulation.}
FastServe uses token-level preemption and a skip-join multilevel feedback queue to reduce head-of-line blocking \citep{wu2026fastserve}. VTC accounts for processed input and output tokens to allocate service fairly across clients under continuous batching \citep{sheng2024fairness}. Llumnix adapts request placement through runtime rescheduling and live migration \citep{llumnix}. Simulation provides a complementary way to evaluate configuration choices: Vidur uses profiling-calibrated models to explore serving configurations \citep{vidur}, while LLMCompass models inference execution and searches mappings to compare hardware designs \citep{zhang2024llmcompass}. These systems advance online scheduling and performance estimation. \sys uses snapshot-based prediction within an online controller for a different objective: admit configurations by predicted deadlines and rank their interval-energy contribution against a supplied commitment.

\section{System Overview}
\label{sec:overview}

\begin{figure}[htbp]
  \centering
  \includegraphics[width=\linewidth]{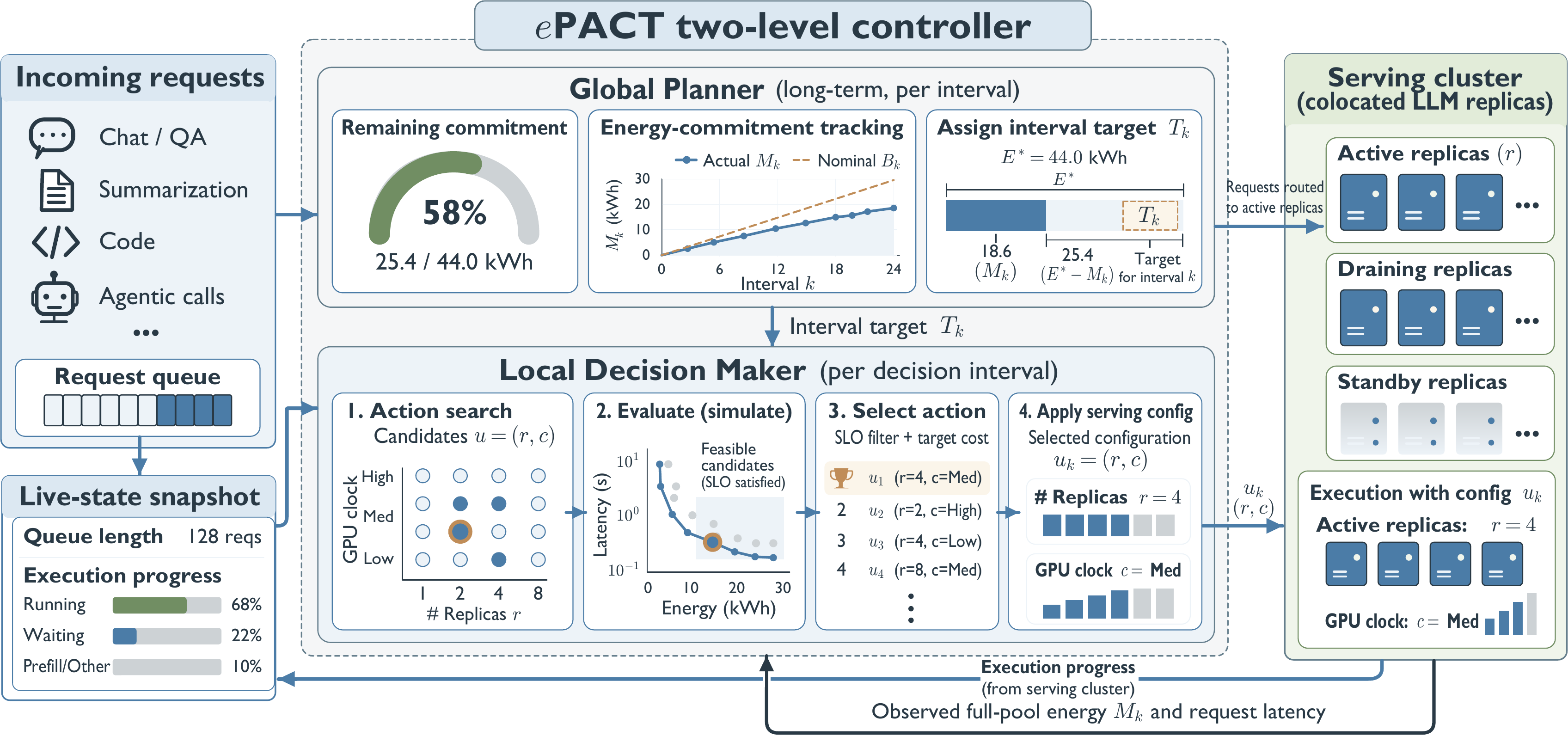}
  \caption{\sys architecture. Remaining commitment, cumulative-energy tracking, and interval-target assignment belong to the global planner. Adaptive search, prediction, SLO checking, selection, and configuration application belong to the local decision maker. The serving pool co-locates prefill and decode. Mini-plots illustrate state and control, not measured results.}
  \label{fig:architecture}
\end{figure}

\sys translates a supplied hourly energy commitment into online serving decisions that seek to reduce two-sided deviation costs while protecting service quality. The global planner updates interval energy targets; the local decision maker searches for, evaluates, and selects configurations for a co-located serving pool. Figure~\ref{fig:architecture} shows the component responsibilities and feedback loop.

\subsection{Control Components}
\label{sec:components}

\noindent\textbf{Global planner.}
The global planner tracks cumulative energy consumption within the current hour to determine the remaining energy commitment. At each decision boundary, it allocates the remaining commitment across the remaining intervals and supplies the current interval target. Recomputing this target from measured consumption allows subsequent decisions to account for earlier overuse or underuse instead of following a fixed allocation.

\noindent\textbf{Local decision maker.}
The local decision maker searches over serving configurations defined by two parameters: the \emph{active-replica count}, which determines how many replicas accept new requests, and the \emph{GPU clock}, which sets their GPU operating frequency. Using the target and a snapshot of current requests and replica states, a profiling-calibrated simulator predicts each candidate's interval energy consumption and request completion times. Among candidates that pass the predicted SLO check, the local decision maker selects the configuration with the lowest asymmetric deviation cost relative to the interval energy target. Section~\ref{sec:method} formalizes target updates, selection, and fallbacks.

\subsection{Serving and Feedback}
\label{sec:serving_feedback}

Arrivals and execution progress update the live state captured at each decision boundary. Serving continues under the current configuration during search; a selected configuration is applied only if it remains valid. A replica removed from active service finishes its running requests before entering standby, and standby replicas retain model weights for reactivation. Energy accounting covers active, draining, and standby replicas, so reducing active capacity does not remove idle consumption from the commitment ledger. These measurements feed the next global target update; Appendix~\ref{app:implementation} details routing and application.

\section{Energy-Commitment Tracking}
\label{sec:method}

The controller must connect an hour-end energy objective to decisions made before future requests are known. We first state that objective, then define feedback-based interval targets and the predictive selection rule used to pursue it online.

\subsection{Problem Formulation}
\label{sec:formulation}

\noindent\textbf{Input.}
Requests arrive online with varying prompt and response lengths. These lengths and arrival fluctuations determine the batches formed by the runtime and the capacity needed to meet deadlines. The controller observes arrived requests and their execution progress, but not future arrivals. For each hour, $\mathcal R$ denotes its arrival cohort. Requests retain their original deadlines across hour boundaries, and their SLO outcomes belong to their arrival hour even when they complete later. Serving state persists across hours.

\noindent\textbf{Energy commitment.}
We consider contracts that assign positive costs to deviations on either side of a committed energy volume. Each hour has a preset commitment $E^*>0$ and rates $\lambda_+>0$ and $\lambda_->0$, measured in cost per unit energy, for overuse and underuse. These are fixed inputs, not procurement decisions made by \sys. The fixed procurement payment is common to the compared policies and omitted; the objective models deviation charges rather than a complete electricity bill. We omit the hour index below and track the controlled GPU pool; Appendix~\ref{app:energy_scope} explains its relationship to facility consumption.

\Needspace{3\baselineskip}
\noindent\textbf{Constraints.}
The service requirement constrains acceptable serving configurations. Request $i$, arriving at time $a_i$, has a reference latency $L_i^{\mathrm{ref}}>0$: its standalone service time on an isolated replica under a fixed reference configuration. This reference is independent of the candidate serving action, so changing the action does not change the request's deadline. The SLO scale $\alpha>0$ defines that deadline as $d_i=a_i+\alpha L_i^{\mathrm{ref}}$. The desired service policy permits a deadline-miss fraction of at most $\epsilon\in[0,1]$ for each arrival cohort; $\epsilon=0$ permits none.

\noindent\textbf{Action space.}
\sys adjusts active-replica count and GPU clock jointly. A configuration $u=(r,c)\in\mathcal U=\{1,\ldots,N\}\times\mathcal C$ activates $r$ replicas from a resident pool of size $N$, where $N$ is a positive integer and $\mathcal C$ is the finite set of supported GPU clock levels. More active replicas provide greater concurrent capacity but also keep more resources active; changing the clock adjusts execution speed and power within each replica. Both choices affect batch formation, request completions, and the backlog carried forward. Deactivated replicas drain running requests before entering standby, and all three states remain inside the energy-accounting boundary.

\noindent\textbf{Optimization objective.}
Let hour start time be $t_0$, and divide its settlement horizon into $K$ intervals with boundaries $t_k=t_0+k\Delta$, where $K$ is a positive integer, $\Delta>0$, and $K\Delta=3{,}600\,\mathrm{s}$. A causal policy $\pi$ initiates a decision from the observed state at each boundary and continues operating across hours. Write $u_k$ for the candidate selected by decision $k$. Search and serving proceed asynchronously, so $u_k$ need not be applied at $t_k$ or remain in force throughout that interval. If $P_{\mathrm{pool}}^\pi(t)$ is actual GPU-pool power under the applied configuration history, interval energy is $e_k=\int_{t_k}^{t_{k+1}}P_{\mathrm{pool}}^\pi(t)\,\mathrm{d}t$ and total consumption is $M_K=\sum_{k=0}^{K-1}e_k$. This accounting includes consumption during planning and configuration transitions. With $f_i$ denoting actual completion time under the continuing policy, the desired objective for the hour is
\begin{equation}
 \begin{aligned}
  \min_{\pi}\quad & C(M_K)=\lambda_+\pos{M_K-E^*}
                         +\lambda_-\pos{E^*-M_K},\\
  \text{subject to}\quad &
  \bigl|\{i\in\mathcal R:f_i>d_i\}\bigr|
  \leq\epsilon|\mathcal R|,
 \end{aligned}
 \label{eq:terminal}
\end{equation}
where $[x]_+=\max(x,0)$. This objective penalizes both directions of energy deviation under the prescribed service requirement. Energy belongs to the hour in which it is consumed, including energy used by requests carried over from earlier hours. Each hour is settled separately, so overuse in one hour is not offset by underuse in another, although the serving dynamics couple successive hours. The controller below pursues this objective using predictions and feedback; we do not claim an optimal online policy or guaranteed feasibility under arbitrary arrivals.

\noindent\textbf{From an hourly commitment to interval targets.}
Equation~\eqref{eq:terminal} assesses energy at the end of each hour, but serving decisions require a near-term reference. \sys therefore assigns each interval an adjustable energy target $T_k$ that guides current decisions toward the remaining commitment. The global planner computes this target from measured progress; the local decision maker uses it to rank predicted configurations. These interval scores are a control heuristic, not an additive decomposition of the terminal cost.

\subsection{Global Planner: Dynamic Interval Targets}
\label{sec:global}

The global planner converts cumulative consumption into the next interval's energy target. Keeping every target fixed would leave earlier overuse or underuse uncorrected; updating targets lets the remaining intervals compensate within the same hour.

\noindent\textbf{Allocating the interval energy target.}
Suppose the operator has a prior nominal allocation $b_k\geq0$ for each decision interval, with $\sum_{k=0}^{K-1}b_k=E^*$ and positive remaining nominal mass $\sum_{j=k}^{K-1}b_j>0$ for every $k<K$. At the start of interval $k$, the planner has observed cumulative consumption $M_k=\sum_{j<k}e_j$, so the signed energy remainder is $E^*-M_k$. Meanwhile, if $B_k=\sum_{j<k}b_j$ denotes the nominal allocation assigned to elapsed intervals, then $E^*-B_k>0$ is the nominal allocation remaining over the rest of the hour.

\noindent\textbf{Scaling the nominal allocation.}
The ratio $\frac{(E^*-M_k)}{(E^*-B_k)}$ compares the remaining commitment with the remaining nominal allocation. A value above one indicates less consumption than planned and raises the target; a value below one indicates more consumption and lowers it. The planner scales $b_k$ accordingly:
$T_k=b_k\frac{E^*-M_k}{E^*-B_k}.$
Scaling by this ratio retains the relative nominal allocations while correcting for measured consumption. By default, the nominal allocations are equal, $b_k=\frac{E^*}{K}$, so the target update becomes
$ T_k=\frac{E^*-M_k}{K-k}.$
The next interval therefore receives an equal share of the actual remaining commitment, not an unchanged share of the original commitment.

\begingroup
\setlength{\fboxsep}{7pt}
\setlength{\fboxrule}{0.6pt}
\definecolor{examplefill}{HTML}{F4F7F3}
\noindent\fcolorbox{black!30}{examplefill}{%
\begin{minipage}{\dimexpr\linewidth-2\fboxsep-2\fboxrule\relax}
\textbf{Illustrative example for determining next interval's energy target.}
Consider an hourly commitment of $E^*=600\,\mathrm{kJ}$ divided into $K=6$ intervals with nominal allocation $b_k=100\,\mathrm{kJ}$ each. After the first interval, $B_1=100\,\mathrm{kJ}$ and five intervals remain. If $M_1=70\,\mathrm{kJ}$, underuse leaves $530\,\mathrm{kJ}$ to allocate, raising the next target to $T_1=100\frac{600-70}{600-100}=106\,\mathrm{kJ}$. If $M_1=130\,\mathrm{kJ}$ instead, overuse leaves $470\,\mathrm{kJ}$, lowering the next target to $T_1=100\frac{600-130}{600-100}=94\,\mathrm{kJ}$. Only the current target is issued; subsequent targets are recomputed from updated measurements rather than fixed in advance.
\end{minipage}}
\par\endgroup

\noindent\textbf{Service priority and target reachability.}
An interval target is a cost reference, not a hard cap. Bursts may require extra energy to meet SLOs, while little useful work can make a high target unreachable. Consumption above $E^*$ can even yield a negative target, not negative physical energy. The local decision maker therefore checks predicted service quality before matching $T_k$.

\subsection{Local Decision Maker: SLO Checks and Target Matching}
\label{sec:local}

Given the interval target, request SLOs, and live-state snapshot, the local decision maker checks predicted service quality before ranking configurations by energy-deviation cost. Appendix Figure~\ref{fig:local_selection} illustrates the choice; Section~\ref{sec:search} explains how adaptive search proposes candidates.

\noindent\textbf{Predicting completion times and energy.}
Every candidate starts from the same captured state. The simulator replays the serving scheduler, substituting profiling-calibrated step duration and power for GPU execution. It also replays previously observed arrivals, without accessing future trace arrivals. Energy prediction $\widehat E_k(u)$ covers a simulated interval of duration $\Delta$ from that snapshot; completion prediction $\widehat f_i(u)$ continues the candidate configuration until the captured requests finish. These horizons differ because a request extending beyond the interval must still be checked against its deadline. The energy prediction is a snapshot-based surrogate for ranking actions. It is distinct from actual metered $e_k$, which includes ongoing service before an asynchronous result is applied. Later decisions and unseen arrivals can likewise change actual completion times. We calibrate and validate the predictor offline; Appendix~\ref{app:eval_validation} reports prediction accuracy and control overhead.

\noindent\textbf{Admitting candidates by SLO.}
Let $\mathcal I_k$ be the unfinished requests in the snapshot, including requests carried over from earlier hours. The controller compares their predicted completion times with their original deadlines, including delay already incurred. Using $\epsilon$ as the local miss tolerance, a candidate passes when its predicted miss count satisfies
\begin{equation}
 \bigl|\{i\in\mathcal I_k:\widehat f_i(u)>d_i\}\bigr|
 \leq\epsilon|\mathcal I_k|.
 \label{eq:gate}
\end{equation}
Let $\mathcal F_k\subseteq\mathcal U$ denote the evaluated candidates with complete, valid predictions that pass this check. With $\epsilon=0$, every captured request must be predicted to finish on time. Equations~\eqref{eq:terminal} and~\eqref{eq:gate} use the same adjustable tolerance on different populations: the first assesses actual outcomes for an arrival cohort, whereas the second assesses predicted outcomes for unfinished requests. Completed misses are absent from $\mathcal I_k$, and future actions can change remaining requests' outcomes. The local gate is therefore a predictive service heuristic, not enforcement of the cohort constraint, even with accurate snapshot predictions. When $\mathcal I_k$ is empty, the miss-count check is vacuously satisfied.

\Needspace{6\baselineskip}
\noindent\textbf{Matching the target among admitted candidates.}
Among admitted candidates, neither lower energy nor lower latency alone determines the choice. Instead, the local decision maker compares each candidate's predicted energy with $T_k$ using the terminal objective's overuse and underuse rates:
\begin{equation}
 \begin{split}
 J_k(u)&=\lambda_+\pos{\widehat E_k(u)-T_k}
              +\lambda_-\pos{T_k-\widehat E_k(u)}, \,\, u_k\in\argmin_{u\in\mathcal F_k}J_k(u).
 \end{split}
 \label{eq:selection}
\end{equation}
The selected candidate is the closest \emph{cost-weighted} match, which need not minimize absolute energy deviation. Thus, passing the SLO check establishes eligibility; the target and penalty rates determine the energy preference. The hour-end objective remains Equation~\eqref{eq:terminal}.

\noindent\textbf{Service-first fallback.}
If no evaluated candidate passes the check, \sys first minimizes predicted deadline misses among candidates with valid predictions. Ties favor the largest worst-request deadline margin, $m_k(u)=\min_{i\in\mathcal I_k}\{d_i-\widehat f_i(u)\}$, measured in units of time, and then the lowest energy-deviation cost. A positive margin places every captured request inside its deadline. Equal-cost admitted candidates use the same service-based tie-breaks. For an empty snapshot, all candidates have zero predicted misses and no request-specific margin preference. If no candidate has a valid prediction, \sys retains the current configuration.

\subsection{Adaptive Action Search}
\label{sec:search}

Local selection can rank only the configurations already evaluated; adaptive search determines which to evaluate next. \sys starts broadly, then refines promising neighborhoods. The search retains evaluated configurations and their predicted request completion times and interval energy, using these records to choose unvisited candidates until the stopping condition below is met.

\noindent\textbf{Constructing the catalog.}
\sys selects up to eight replica counts and five supported GPU clocks at approximately uniform rank intervals, preserving both endpoints of each domain, yielding at most $A=40$ configurations. The search assumes no monotonic relationship between replica count, clock frequency, energy, or SLO attainment.

\noindent\textbf{Coarse initialization and warm starts.}
Each search begins with a coarse lattice formed from the lowest, middle, and highest ranks along both dimensions, augmented with the current configuration and historical warm-start candidates. Duplicate configurations are removed. Historical configurations seed the search, but their predictions and scores are recomputed for the current decision.

\noindent\textbf{Adaptive refinement.}
Starting from the initial candidates, \sys performs a coarse-to-fine search around the best configuration found so far, where ``best'' follows the selection rule in Section~\ref{sec:local}. At each resolution, it evaluates unvisited neighboring configurations by changing one or both configuration dimensions by the current search stride. The search starts with a relatively coarse stride, approximately one quarter of each dimension's rank span. Whenever a better configuration is found, \sys recenters the search at the same stride. Once all neighbors at the current resolution have been explored, it halves the stride and continues at a finer resolution. The search stops after all one-step neighbors of the best configuration have been evaluated. If none of the evaluated configurations has a complete, valid prediction, \sys evaluates the remaining configurations as a fallback.

\noindent\textbf{Execution and overhead.}
Candidates proposed together are evaluated in parallel using persistent CPU workers. Search runs asynchronously with serving and terminates by completion of the unit-step neighborhood or catalog exhaustion, without a wall-time cutoff. If $n_{\mathrm{eval}} \leq A$ configurations are evaluated, the simulation work is $O(n_{\mathrm{eval}}T_{\mathrm{sim}})$, where $T_{\mathrm{sim}}$ denotes the cost of one candidate simulation. Each configuration is evaluated at most once per decision.

\FloatBarrier
\section{Evaluation}
\label{sec:evaluation}

We evaluate whether \sys reduces hourly commitment-deviation costs while preserving request-level service quality, focusing on four questions:
\begin{itemize}[leftmargin=*]
 \item \textbf{Q1: Single-hour service-energy performance.} How does \sys compare with baselines across SLO scales under fixed and decreasing commitments?
 \item \textbf{Q2: Controller ablation.} How do the global planner and local decision maker in \sys contribute to SLO attainment and hourly commitment tracking?
 \item \textbf{Q3: Full-day commitment tracking.} At a fixed SLO scale, how does \sys perform over a full day of changing workloads and hourly commitments?
 \item \textbf{Q4: Prediction and planning.} How accurate are the measured predictions, and what planning time does the controller incur?
\end{itemize}
The single-hour study tests sensitivity to SLO and commitment choices and characterizes the controller components. The full-day study evaluates continuous tracking, while held-out measurements assess the predictions used for local selection.

\subsection{Experimental Setup}
\label{sec:setup}

\noindent\textbf{Serving setups.}
We study H20 and H200 serving setups, each with a 32-GPU pool and one resident vLLM-hosted single-GPU replica of Qwen3.8-27B-FP8 per device. H20 and H200 single-hour studies use physical executions; both 24-hour studies use CPU simulation with GPU-specific profiles. Each hourly commitment spans 60 one-minute control intervals. Energy covers the controlled GPU pool, including active, draining, and standby consumption. Appendices~\ref{app:eval_protocol} and~\ref{app:energy_scope} detail execution and the device-energy measurement boundary.

\noindent\textbf{Workloads.}
Both studies use requests from the Azure LLM inference trace \citep{azure2024}; the full-day study replays $969{,}781$ requests.

\noindent\textbf{Preset hourly commitments.}
The single-hour study uses two settings. A \emph{fixed commitment} tests sensitivity to the SLO requirement without changing the energy objective. A \emph{decreasing commitment} tests whether looser SLOs allow tracking lower energy targets. In both settings, a run's hourly commitment is fixed; ``decreasing'' describes the targets assigned across SLO scales, not a change within a run. Full-day commitments instead vary by hour. For H20, each target scales vLLM's hourly energy on a separate historical day matching the evaluation day's weekday; the fraction is calibrated using other configurations that meet the historical workload's SLO-attainment requirement. H200 uses the same fractions with a GPU-specific offline evaluation-day vLLM reference. All targets are frozen before the compared controller runs and shared across methods; Appendix~\ref{app:eval_calibration} gives their values and construction.


\noindent\textbf{Baselines.} We compare \sys against two representative baselines:
\begin{itemize}[leftmargin=*]
    \item \noindent\textbf{vLLM.}
We use unmodified vLLM as the latency-first representative, keeping all 32 replicas resident and active throughout each run. GPUs can operate up to the experiment's hardware-specific clock ceiling; this baseline does not proactively reduce clocks or deactivate replicas to save energy.
\item \noindent\textbf{DynamoLLM.}
We reimplement DynamoLLM's energy-aware serving policy as the energy-minimizing representative. This evaluated policy adjusts active capacity and GPU clocks from observed demand and latency pressure, and restores capacity for urgent requests. It uses the same model, workload, and GPU pool as \sys.
\end{itemize}

\noindent\textbf{Ablations.} We evaluate two variants that successively remove the controller's two levels:
\begin{itemize}[leftmargin=*]
    \item \textbf{Static-Target.}
    This variant removes the global planner and uses fixed nominal targets, $T_k=b_k$, without correcting accumulated energy deviation. The local decision maker still checks predicted SLOs and adjusts replica counts and GPU clocks to minimize deviation cost.
    \item \textbf{Fixed-Config.}
    This variant also removes the local decision maker and keeps one configuration throughout the hour: 28 active replicas at 1980\,MHz on H20 and 1710\,MHz on H200.
\end{itemize}


\noindent\textbf{Metrics.} We assess service quality and commitment tracking together:
\begin{itemize}[leftmargin=*]
    \item \noindent\textbf{SLO attainment.}
We report the fraction of requests meeting the deadlines in Section~\ref{sec:formulation}. Single-hour comparisons and ablations use $\alpha\in\{2.5,3,3.5\}$; the full-day comparison fixes $\alpha=3$. Requests are grouped by arrival time and scored at completion, including completions after the evaluation horizon. Full-day attainment pools all arrivals rather than averaging hourly percentages.
\item \noindent\textbf{Commitment cost and tracking error.}
For reporting, overuse and underuse receive dimensionless weights of 2 and 1, respectively; the same weights apply to the single-hour cost comparison below. If $E_h$ and $E_h^*$ are consumed and committed GPU energy in kWh, the full-day score is $\sum_{h=0}^{23}\bigl(2[E_h-E_h^*]_+ + [E_h^*-E_h]_+\bigr)$, in kWh-equivalent. This measures asymmetric deviation exposure under a fixed experimental rate ratio, rather than an electricity bill. Each hour settles separately, so opposite deviations cannot cancel. Signed hourly deviation is $\delta_h=100(E_h-E_h^*)/E_h^*$, expressed as a percentage; positive values indicate overuse. We report the unweighted mean absolute hourly deviation $\frac{1}{24}\sum_{h=0}^{23}|\delta_h|$ and the number of hours with $|\delta_h|\leq5$. Energy is assigned to the hour in which it is consumed. Appendix~\ref{app:eval_measurement} relates the score to the monetary objective.
\end{itemize}

\subsection{Single-Hour Service-Energy Comparison}
\label{sec:mainresults}

\begin{figure}[htbp]
 \centering
 \includegraphics[width=\linewidth]{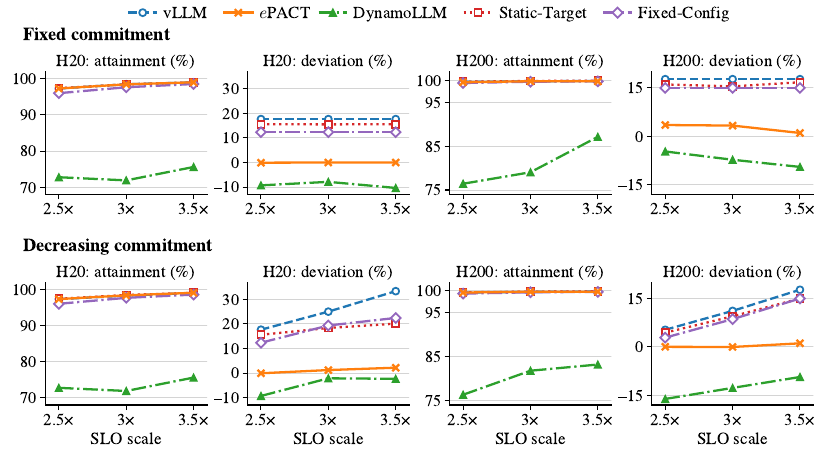}
 \caption{Single-hour comparisons with fixed (top) and decreasing (bottom) commitments. Columns show H20 and H200 attainment and signed deviation; positive deviation denotes overuse.}
 \label{fig:single_hour_main}
 \vspace{-1em}
\end{figure}

Figure~\ref{fig:single_hour_main} answers Q1 for both settings on H20 and H200. We discuss baselines first, then ablations in Section~\ref{sec:ablation_eval}. Appendix Tables~\ref{tab:h20_single_hour_appendix} and~\ref{tab:h200_single_hour_appendix} list the values.

\noindent\textbf{\sys keeps high attainment while aligning to the preset commitment.}
Across both settings and GPU types, \sys stays within $0.09$ percentage points of vLLM's attainment and achieves the smallest absolute deviation among the five methods. With fixed commitments, absolute deviation is at most $0.09\%$ on H20 and $3.46\%$ on H200, preserving tracking as the SLO changes. With decreasing commitments, the corresponding maximum deviations are $2.18\%$ and $1.08\%$: \sys uses the flexibility of looser SLOs to track lower targets. H200 energy falls from $15.27$ to $13.81$ MJ while attainment remains above $99.63\%$.

\noindent\textbf{Full-capacity serving does not track commitments.}
vLLM overuses the fixed commitments by $17.65\%$ on both GPUs. Under decreasing commitments, its overuse rises from $17.65\%$ to $33.33\%$ on H20 and from $5.26\%$ to $17.65\%$ on H200. The increase reflects smaller target denominators, not increased serving capacity: the baseline has no feedback that adapts consumption to the commitment.

\noindent\textbf{Lower consumption alone does not establish a better service-cost outcome.}
Across both settings, the evaluated DynamoLLM policy undershoots the commitments and has lower SLO attainment: $71.88\%$-$75.59\%$ on H20 and $76.34\%$-$87.12\%$ on H200. These runs show that reducing consumption alone does not ensure commitment alignment or preserve service quality. Absolute deviation and asymmetric cost must also be distinguished: at H20 $3.5\times$ with a decreasing commitment, \sys's $+2.18\%$ deviation incurs more weighted cost than DynamoLLM's $-2.34\%$, but attainment is $99.06\%$ versus $75.59\%$. Thus, the comparison supports a joint service-and-tracking benefit, not uniformly lower cost regardless of service quality.

\subsection{Controller Ablation}
\label{sec:ablation_eval}

To answer Q2, we compare \sys with Static-Target and Fixed-Config within each setting in Figure~\ref{fig:single_hour_main}.

\noindent\textbf{Global feedback corrects accumulated deviation.}
Removing the global planner increases absolute deviation in every comparison, while Static-Target's attainment remains within $0.13$ percentage points of \sys. Its overuse is $15.41\%$-$16.69\%$ with fixed commitments and $4.37\%$-$20.11\%$ with decreasing commitments. For H200 at $3\times$ under the latter setting, global feedback reduces deviation from $+9.43\%$ to $-0.05\%$. Feedback corrects the cumulative error left by fixed targets.

\noindent\textbf{Local selection preserves flexibility.}
With the global planner disabled, Static-Target still adapts configurations to minimize predicted deviation cost subject to the SLO check; Fixed-Config cannot adapt. Static-Target has higher SLO attainment throughout both settings. Fixed-Config often has lower energy deviation, but sacrifices attainment: at H20 $2.5\times$, the gap is $1.38$ percentage points. With decreasing H20 commitments at $3\times$ and $3.5\times$, local selection improves both metrics, reducing overuse to $18.35\%$ and $20.11\%$ from $19.36\%$ and $27.32\%$. Local selection therefore improves service quality relative to the evaluated fixed configuration, while its tracking benefit depends on the setting. These nested ablations characterize complementary roles: local adaptation responds to the workload, and global feedback corrects accumulated energy deviation.

\subsection{24-Hour Commitment Tracking}
\label{sec:tracking_eval}

To answer Q3, Table~\ref{tab:tracking_results} compares \sys with the baselines in 24-hour simulations at fixed $\alpha=3$ using H20 and H200 profiles. Unlike the single-hour study, both arrivals and commitments change across hours. Requests and serving state persist across hour boundaries, as detailed in Appendix~\ref{app:eval_measurement}.

\begin{table}[htbp]
 \centering
 \caption{Full-day simulation at fixed $3\times$ SLO. Energy metrics use 24 separate hourly settlements; attainment pools all arriving requests. The hour count is out of 24. Cost uses the experimental 2:1 overuse-to-underuse weights and is reported in kWh-equivalent.}
 \label{tab:tracking_results}
 \small
 \setlength{\tabcolsep}{4.5pt}
 \begin{tabular*}{\linewidth}{@{\extracolsep{\fill}}llrrrr@{}}
  \toprule
  GPU & Method & Attainment (\%) & Mean abs. dev. (\%) & Hours within $\pm5\%$ & Cost \\
  \midrule
  H20 & vLLM & 99.4836 & 11.4304 & 7 & 23.5683 \\
  H20 & DynamoLLM & 80.2132 & 8.9597 & 5 & 14.3261 \\
  H20 & \sys & \textbf{99.4892} & \textbf{2.1630} & \textbf{22} & \textbf{6.1838} \\
  \midrule
  H200 & vLLM & 99.9644 & 11.4304 & 7 & 18.6780 \\
  H200 & DynamoLLM & 83.8908 & 12.5358 & 4 & 15.0900 \\
  H200 & \sys & \textbf{99.9371} & \textbf{2.3092} & \textbf{21} & \textbf{4.5320} \\
  \bottomrule
 \end{tabular*}
\end{table}

\noindent\textbf{Lower daily deviation cost with near-vLLM service quality.}
In the H20 and H200 simulations, \sys reduces the daily deviation cost relative to vLLM by $73.76\%$ and $75.74\%$, respectively. Its SLO attainment is $99.4892\%$ and $99.9371\%$, within $0.03$ percentage points of vLLM in both cases. It also lowers cost relative to the evaluated DynamoLLM policy by $56.84\%$ and $69.97\%$, while attaining substantially more request deadlines. The results support the objective of reducing commitment-deviation exposure while preserving service quality under the evaluated rate ratio.

\noindent\textbf{Consistent tracking across separate hourly settlements.}
Mean absolute hourly deviation falls from vLLM's $11.4304\%$ to $2.1630\%$ on H20 and $2.3092\%$ on H200. Respectively, 22 and 21 of 24 hours fall within $\pm5\%$ of their commitments. These metrics show that the lower aggregate cost accompanies small deviations in most individual hours.

\noindent\textbf{Hourly results reveal the remaining tracking error.}
On H200, \sys exceeds the commitment by $5.80\%$ in hour 7 and $5.26\%$ in hours 21 and 23, despite attainment above $99.9\%$ in each. These residuals distinguish average tracking quality from exact agreement in every hour. A daily energy total could hide offsetting hourly errors; mean absolute deviation measures their magnitude, while the within-band count indicates how consistently the controller tracks separate hourly commitments. Appendix Figures~\ref{fig:h20_day_overview} and~\ref{fig:h200_day_overview} and Tables~\ref{tab:hourly_tracking} and~\ref{tab:h200_hourly_tracking} show the hourly trajectories and outcomes.

\subsection{Prediction Accuracy and Planning Overhead}
\label{sec:prediction_results}

To answer Q4, we examine held-out prediction measurements and the observed time required for asynchronous planning.

\noindent\textbf{Held-out predictor validation.}
We validate the predictions used for local selection on held-out workloads, with detailed results in Appendix~\ref{app:eval_validation}. With frozen models and no refitting, five fixed-batch workload classes, each repeated twice on one replica at 1980\,MHz, yield latency WAPE of $0.44\%$-$2.66\%$ and energy WAPE of $0.19\%$-$2.89\%$. Here WAPE is $100\frac{\sum_j|\widehat y_j-y_j|}{\sum_j y_j}$ for measured values $y_j$ and predictions $\widehat y_j$. Another nonoverlapping trace slice uses 97 requests with continuous arrivals and batching. Latency WAPE is $4.15\%$, while the 95th percentile of per-request relative error is $14.74\%$ and 84 of 97 predictions fall within $\pm5\%$ of measured latency. The signed energy error over the 120-second window is $+0.76\%$; this window error is distinct from energy WAPE. These checks characterize the evaluated single-replica workloads. The larger tail of prediction errors also shows why low aggregate error does not ensure every request-level SLO prediction is accurate.

\noindent\textbf{Asynchronous planning cost.}
The reported run completes 60 asynchronous decisions with a mean search wall time of $23.35$ seconds per decision, compared with a 60-second control interval. Serving continues during search, and the controller checks whether a result remains valid before applying it. The measurement establishes the observed planning time; without a search baseline or latency-impact comparison, it does not quantify a search speedup or establish zero serving overhead.

\FloatBarrier
\section{Conclusion}
\label{sec:conclusion}

We presented \sys, an LLM serving system for preset hourly energy commitments with positive two-sided deviation costs. In this regime, energy reduction alone is insufficient: serving must account for contractual deviation exposure and service quality together. \sys combines measured-energy feedback with predictive configuration selection to pursue this objective online. In 24-hour trace simulations for H20 and H200 pools, it reduces asymmetric deviation cost by $73.8\%$ and $75.7\%$ relative to vLLM, with near-vLLM attainment and mean absolute hourly deviation of at most $2.31\%$. These results show that adapting serving choices to measured progress toward a commitment can preserve service quality while reducing modeled deviation exposure.

\label{main-text-end}

\clearpage
\section*{AI use statement}

In this work, we used generative AI tools to aid and polish writing, including language refinement, grammar checking, typo correction, and improving clarity and readability. 
We did not use generative AI tools to generate research ideas, design the methodology, or conduct experiments. 
All AI-assisted text was reviewed and revised by the authors. 
We take responsibility for the final content of this work, including text, claims, and artifacts produced with the aid of generative AI.
\bibliographystyle{plainnat}
\bibliography{references}

\ifappendixdetails
\clearpage
\appendix
\section{Additional Implementation Details}
\label{app:implementation}

This appendix details replica management, prediction semantics, the measurement boundary, and supplementary evaluation results.

\noindent\textbf{Component boundaries and feedback.}
Figures~\ref{fig:planner} and~\ref{fig:local_selection} detail the two control components shown in Figure~\ref{fig:architecture}. The hourly commitment is supplied externally. Measured GPU energy updates the global planner; arrivals and execution progress update the live state captured by the local decision maker.

\begin{figure}[htbp]
  \centering
  \includegraphics[width=\linewidth]{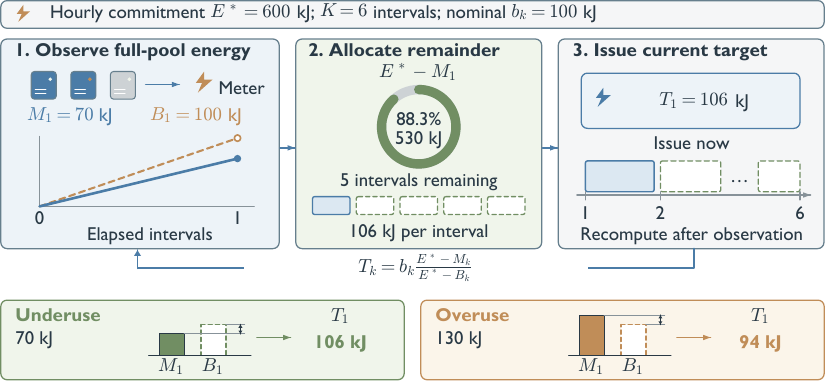}
  \caption{Global planning from measured progress to the current interval target. Dashed energy bars mark 100\,kJ. Timeline labels denote decision boundaries, with $K=6$ the hour end; five intervals remain after boundary 1. Future targets are recomputed after observation.}
  \label{fig:planner}
\end{figure}

\begin{figure}[htbp]
  \centering
  \includegraphics[width=\linewidth]{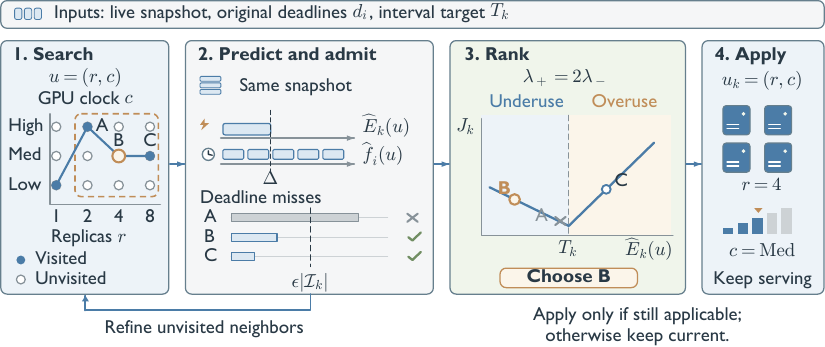}
  \caption{Local prediction and action selection. A fails the SLO check; B has lower weighted deviation cost than C despite being farther from the energy target.}
  \label{fig:local_selection}
\end{figure}

\noindent\textbf{Autoscaling and replica lifecycle.}
The lifecycle in Section~\ref{sec:serving_feedback} preserves in-flight execution during capacity changes. A replica selected for deactivation stops accepting new requests and drains its running requests; reactivation uses model weights retained in GPU memory rather than reloading them.

\noindent\textbf{Request routing.}
The router assigns each incoming request to an active replica with the fewest unfinished requests. When the active-replica count changes, queued requests that have not begun execution may be redistributed among active replicas. Running requests remain on their current replicas, including those draining toward standby.

\noindent\textbf{Asynchronous decisions.}
Each control decision uses a snapshot captured at the decision boundary. Serving continues under the applied configuration while search evaluates hypothetical candidates. A returned candidate is applied only if it remains valid for the live state; otherwise, the current configuration is retained. Thus, a snapshot's one-interval energy prediction is a ranking surrogate, not a record of energy actually consumed after the snapshot. The measured ledger includes execution during planning, delayed application, and replica transitions. Hourly boundaries reset that ledger and the commitment, without resetting queued requests or their deadlines. The reported mean planning time does not establish a bound on application delay or the frequency of stale decisions.

\lead{Native-step projection}
Beyond the projection described in Section~\ref{sec:local}, the simulator preserves queue and cache semantics, progress feedback, and pending steps; a pending step completes before a clock transition. The current calibration combines phase-specific duration components, a decode plateau model, and empirical power corrections. A different GPU family requires new calibration of these numerical components; the controller and search alone do not establish prediction accuracy.

\section{Scope and Limitations}
\label{app:limits}
Our economic premise is that reducing physical energy consumption alone does not guarantee lower deployment cost when service obligations and contractual deviation charges also matter. Equation~\eqref{eq:terminal} explicitly assumes positive costs for both overuse and underuse. This is a model of two-sided deviation exposure, not a consequence of every advance purchase. PJM establishes financially binding day-ahead positions and separately settles real-time differences \citep{pjm_billing}. Surplus settlement can provide a credit; EDF's volume-tolerance example instead illustrates conditional charges associated with unused contracted energy. WAPA's DSW-EI4 schedule provides a transmission-level example of deficit premiums and discounted surplus credits. These arrangements differ in settlement periods, tolerance bands, and netting rules, which our hourly linear model does not reproduce. A sunk procurement payment or resale below its acquisition price alone does not imply a positive marginal underuse charge. The modeled charge must represent the applicable contractual exposure. Fixed procurement payments are common across the compared policies and excluded; the resulting score is not a complete electricity bill. The experimental weights in Appendix~\ref{app:eval_measurement} do not represent a universal market ratio.

The predicted SLO check does not guarantee actual deadlines or target reachability: standby power creates a consumption floor, insufficient work can prevent reaching a high commitment, and heavy load can require more energy than committed. \sys accelerates useful requests rather than generating work to consume energy. Energy storage can complement serving control, but requires a separate dispatch policy and has capacity and conversion-efficiency limits \citep{pjm_storage}. Procurement, storage dispatch, and tariff-band optimization are outside the evaluated controller.

\subsection{GPU Energy and Facility-Level Commitments}
\label{app:energy_scope}

\noindent\textbf{What the energy metric includes.}
We track the electrical energy of the controlled GPU pool, including associated GPU memory and active, draining, and standby consumption. This is device energy, not CPU, host-memory, networking, cooling, or total facility energy. NVIDIA's Management Library (NVML), also used by \texttt{nvidia-smi}, exposes device power and cumulative energy counters \citep{nvidia_nvml,nvidia_smi}. The implemented measurement path integrates timestamped power observations over the accounting interval and sums energy across the pool; it retains cumulative counters for continuity checks. These readings include GPU-associated circuitry and are not per-request or per-process energy measurements. The 24-hour simulator estimates the same device-level quantity from calibrated power and execution duration rather than a facility meter.

\noindent\textbf{Mapping a facility commitment to a GPU target.}
Power usage effectiveness (PUE) is total facility energy divided by total IT-equipment energy; its denominator includes servers, storage, and networking, not only GPUs \citep{greengrid_pue}. For deployment, let $F_h^*$ be the facility commitment in hour $h$, $\widehat\rho_h$ an expected PUE for that hour, and $\widehat E_h^{\mathrm{other\,IT}}$ the expected IT energy outside the controlled GPU pool. An operator can allocate the residual GPU commitment as
\begin{equation}
 E_h^{*,\mathrm{GPU}}=\frac{F_h^*}{\widehat\rho_h}-\widehat E_h^{\mathrm{other\,IT}}.
 \label{eq:facility_mapping}
\end{equation}
A positive residual becomes the controller's supplied $E^*$; a nonpositive residual requires revisiting the facility allocation. Dividing by PUE alone gives an IT budget and approximates a GPU budget only when other IT consumption is negligible or has already been budgeted separately. This mapping lets \sys control the serving-dependent component of a facility commitment without treating GPU telemetry as a utility meter.

\noindent\textbf{Operational scope of the mapping.}
Expected PUE and other IT demand should be estimated for the relevant load and hour, then checked against facility measurements; an annual average PUE is not an exact hourly conversion. GPU tracking translates to facility alignment only to the extent that these estimates hold. The experiments report GPU-pool results without a PUE multiplier and do not claim measured facility-level tracking.

\section{Evaluation Protocol and Supplementary Results}
\label{app:evaluation}

This appendix specifies execution procedures, commitment calibration, and supplementary measurements for Section~\ref{sec:evaluation}.

\subsection{Execution and Configuration}
\label{app:eval_protocol}

\noindent\textbf{Execution and validation scope.}
The single-hour comparisons use matched workloads, SLO definitions, and commitments across the five policies. The 24-hour studies use CPU simulation with GPU-specific profiles. Policy comparisons do not independently establish prediction accuracy; Appendix~\ref{app:eval_validation} reports the scope of the available single-replica validation.

\noindent\textbf{Hardware and runtime.}
The model fits on one GPU, so all methods use tensor-parallel degree one. Beyond the shared setup in Section~\ref{sec:setup}, vLLM~0.27.1 uses a 16{,}384-token context limit, a 2{,}048-token scheduling budget, and at most 64 sequences per replica. H20 clock levels are $\{345,750,1170,1575,1980\}$\,MHz. On H200, we cap the maximum evaluated clock at 1710\,MHz because preliminary runs at 1980\,MHz triggered thermal throttling; Fixed-Config therefore uses 1710\,MHz on H200. One Fixed-Config execution trajectory is scored under the three SLO settings on each GPU setup.

\subsection{Commitment Settings and Calibration}
\label{app:eval_calibration}

Commitment construction occurs \emph{before} online tracking. It answers how much energy to reserve for an expected workload; the controller then tries to track that fixed reservation as the evaluation workload arrives. We distinguish the historical reference, selection of a reduction factor, and application of the resulting target.

\noindent\textbf{Step 1: establish a historical reference.}
For the full-day H20 comparison, the preceding day with the same weekday provides the historical workload: the preceding Saturday is used before replaying Saturday, May~11,~2024. The assumption is that the same weekday has a similar hourly demand pattern, not that the two days have identical arrivals. Let $E_h^{\mathrm{hist}}$ denote the GPU-pool energy used by ordinary full-capacity vLLM on that historical workload in hour $h$. This reference represents operation before applying \sys.

\noindent\textbf{Step 2: calibrate the reduction factor on historical work.}
We evaluate additional serving configurations on the historical workload and identify the lowest measured GPU-pool energy among those meeting the prescribed SLO-attainment requirement for each hour. This empirical estimate guides the hourly fraction $\gamma_h\in(0,1]$ retained from the vLLM reference. Thus, a fraction such as $0.8$ is informed by the observed energy--SLO trade-off on historical requests, rather than chosen solely as an arbitrary savings goal. The calibration gives a minimum over evaluated configurations, not a global lower bound or a guarantee for a different day's workload.

\noindent\textbf{Step 3: freeze the commitment before evaluation.}
The H20 target for evaluation hour $h$ is
\begin{equation}
 E_h^*=\gamma_h E_h^{\mathrm{hist}}.
 \label{eq:historical_commitment}
\end{equation}
For example, a historical reference of $20$\,MJ and $\gamma_h=0.8$ would reserve $16$\,MJ for the corresponding evaluation hour. This illustrative calculation expresses the intention to serve a similar workload using $80\%$ of the historical energy. It does not reveal future arrivals to the online controller. All compared policies receive the same preset targets; \sys adjusts interval targets and configurations, but never revises the hourly commitment after seeing evaluation-day demand. Workload changes can make exact tracking infeasible.

\noindent\textbf{Hourly fractions and the H200 comparison.}
The selected fractions range from $75\%$ to $100\%$. Hours with less calibration headroom use $95\%$ or $100\%$, while less demanding hours allow smaller fractions. Table~\ref{tab:hourly_tracking} lists the H20 fractions and resulting commitments together with evaluation outcomes. The H200 comparison reuses these historical fractions but scales each by the corresponding evaluation-hour vLLM energy from an offline H200 simulation. Its absolute targets are also frozen before the compared controller runs. This separate construction tests tracking supplied targets on the H200 profile; it does not test unseen-day procurement forecasting. Table~\ref{tab:h200_hourly_tracking} reports those targets and outcomes. In both studies, online decisions observe arrived requests and measured progress, not future trace arrivals.

\noindent\textbf{Single-hour target metadata.}
Both settings use SLO scales $2.5$, $3$, and $3.5$. The fixed-commitment setting uses $85\%$ of the corresponding full-capacity vLLM energy at every scale: $16.138982$ MJ on H20 and $13.663689$ MJ on H200. The decreasing-commitment setting uses $85\%$, $80\%$, and $75\%$ on H20, and $95\%$, $90\%$, and $85\%$ on H200. The latter H200 targets are $15.271181$, $14.467435$, and $13.663689$ MJ, respectively. Each target is fixed throughout its run, not decreased online as requests arrive. The two settings compare policies under different prescribed objectives, rather than estimating a minimum feasible energy budget.

\noindent\textbf{Runs for the two single-hour settings.}
The H20 decreasing-commitment results use their corresponding controller runs rather than re-settling the fixed-commitment controller runs. Fixed-Config uses one execution trajectory per GPU setup, with deadlines and deviations rescored for each SLO scale and target. For H20, the decreasing-target deviations of $12.34\%$, $19.36\%$, and $27.32\%$ therefore correspond to the same consumed energy up to rounding. In the decreasing-commitment H200 setting, vLLM (32 active replicas at 1710\,MHz) also uses one execution trajectory scored at the three SLO scales. \sys, Static-Target, and DynamoLLM are executed separately at each scale. The H200 fixed- and decreasing-commitment controller runs are distinct even where their target values coincide.

\subsection{Measurement and Reporting}
\label{app:eval_measurement}

\noindent\textbf{Matched execution and metering.}
Comparisons use the same workload, resident capacity, clock catalog, deadlines, commitments, and settlement boundaries. The GPU measurement boundary is defined in Appendix~\ref{app:energy_scope}. For the single-hour study, SLO attainment includes all arrivals and their full completion tails, whereas energy settlement covers only the first 3{,}600 seconds. Post-hour energy is excluded from hourly deviation and accounted for separately. Single-run attainment differences do not establish statistical significance.

\noindent\textbf{Continuous tracking across hourly boundaries.}
In the full-day replay, queued and running requests, original deadlines, and serving state persist across hourly boundaries; only the commitment and its ledger change. A request's SLO outcome belongs to its arrival hour, even if it completes later, while energy belongs to the hour in which it is consumed. Hours 0--23 are trace-relative indices. The replay is continuous, not 24 independently initialized runs.

\noindent\textbf{Hourly deviation costs.}
We report a nonmonetary, energy-equivalent deviation cost. Let $E_h$ and $E_h^*$ be actual GPU-pool energy and its commitment in hour $h$, both in kWh. Dimensionless weights of 2 for overuse and 1 for underuse give
\begin{equation}
 C_h=2\max(E_h-E_h^*,0)+\max(E_h^*-E_h,0),
 \label{eq:eval_hourly_cost}
\end{equation}
The daily score $\sum_{h=0}^{23}C_h$ has units of kWh-equivalent: it is weighted deviation, not physical energy consumption or a currency charge. The weights express an experimental asymmetry without assuming an electricity price. Equivalently, this score is the monetary objective in Equation~\eqref{eq:terminal} divided by a common reference price per kWh, with a fixed overuse-to-underuse rate ratio. Multiplying by that reference price would restore monetary units; actual tariffs may require different weights or a different settlement function. No fixed procurement payment is added, and deviations in different hours do not cancel.

\noindent\textbf{Daily energy totals.}
For the H20 24-hour study, commitments total $496.1070$~MJ, while vLLM, \sys, and the evaluated DynamoLLM implementation consume $538.5299$, $506.9002$, and $453.9654$~MJ, respectively. DynamoLLM consumes the least energy, but has a larger hourly deviation score and lower SLO attainment than \sys. This comparison illustrates why consumption alone does not determine performance under the modeled objective. Table~\ref{tab:tracking_results} sums hourly costs rather than using a net daily difference; opposite deviations cannot cancel. Hourly entries are rounded independently and need not sum exactly to the full-precision daily totals.

\subsection{Supplementary Figures and Tables}
\label{app:eval_tables}

Tables~\ref{tab:h20_single_hour_appendix} and~\ref{tab:h200_single_hour_appendix} report rounded single-hour values for Figure~\ref{fig:single_hour_main}. Figures~\ref{fig:h20_day_overview} and~\ref{fig:h200_day_overview}, together with Tables~\ref{tab:hourly_tracking} and~\ref{tab:h200_hourly_tracking}, provide the detailed 24-hour views for H20 and H200.

\begin{table}[!htbp]
\centering
\caption{Single-hour H20 results under both commitment settings. Percentages are rounded; deviation is signed. Bold highlights \sys, not column maxima.}
\label{tab:h20_single_hour_appendix}
\small
\setlength{\tabcolsep}{4.5pt}
\begin{tabular*}{\linewidth}{@{\extracolsep{\fill}}lrrrrrr@{}}
\toprule
& \multicolumn{2}{c}{$2.5\times$ SLO} & \multicolumn{2}{c}{$3\times$ SLO} & \multicolumn{2}{c}{$3.5\times$ SLO} \\
Method & Attain. (\%) & Dev. (\%) & Attain. (\%) & Dev. (\%) & Attain. (\%) & Dev. (\%) \\
\midrule
\multicolumn{7}{l}{\textit{Fixed commitment}} \\
vLLM & 97.30 & $+17.65$ & 98.49 & $+17.65$ & 99.00 & $+17.65$ \\
\sys & \textbf{97.31} & \textbf{$-0.09$} & \textbf{98.46} & \textbf{$-0.02$} & \textbf{99.02} & \textbf{$-0.02$} \\
DynamoLLM & 72.75 & $-9.32$ & 71.88 & $-7.89$ & 75.59 & $-10.38$ \\
Static-Target & 97.39 & $+15.58$ & 98.45 & $+15.49$ & 98.97 & $+15.57$ \\
Fixed-Config & 96.01 & $+12.34$ & 97.66 & $+12.34$ & 98.55 & $+12.34$ \\
\midrule
\multicolumn{7}{l}{\textit{Decreasing commitment}} \\
vLLM & 97.30 & $+17.65$ & 98.28 & $+25.00$ & 99.00 & $+33.33$ \\
\sys & \textbf{97.31} & \textbf{$-0.09$} & \textbf{98.34} & \textbf{$+1.22$} & \textbf{99.06} & \textbf{$+2.18$} \\
DynamoLLM & 72.75 & $-9.32$ & 71.88 & $-2.14$ & 75.59 & $-2.34$ \\
Static-Target & 97.39 & $+15.58$ & 98.45 & $+18.35$ & 99.05 & $+20.11$ \\
Fixed-Config & 96.01 & $+12.34$ & 97.66 & $+19.36$ & 98.55 & $+27.32$ \\
\bottomrule
\end{tabular*}
\end{table}

\begin{table}[!htbp]
\centering
\caption{Single-hour H200 results under both commitment settings. Percentages are rounded; deviation is signed. Bold highlights \sys, not column maxima.}
\label{tab:h200_single_hour_appendix}
\small
\setlength{\tabcolsep}{4.5pt}
\begin{tabular*}{\linewidth}{@{\extracolsep{\fill}}lrrrrrr@{}}
\toprule
& \multicolumn{2}{c}{$2.5\times$ SLO} & \multicolumn{2}{c}{$3\times$ SLO} & \multicolumn{2}{c}{$3.5\times$ SLO} \\
Method & Attain. (\%) & Dev. (\%) & Attain. (\%) & Dev. (\%) & Attain. (\%) & Dev. (\%) \\
\midrule
\multicolumn{7}{l}{\textit{Fixed commitment}} \\
vLLM & 99.64 & $+17.65$ & 99.77 & $+17.65$ & 99.86 & $+17.65$ \\
\sys & \textbf{99.57} & \textbf{$+3.46$} & \textbf{99.79} & \textbf{$+3.29$} & \textbf{99.77} & \textbf{$+1.00$} \\
DynamoLLM & 76.43 & $-4.72$ & 79.05 & $-7.28$ & 87.12 & $-9.50$ \\
Static-Target & 99.67 & $+16.01$ & 99.86 & $+15.41$ & 99.89 & $+16.69$ \\
Fixed-Config & 99.39 & $+14.94$ & 99.68 & $+14.94$ & 99.78 & $+14.94$ \\
\midrule
\multicolumn{7}{l}{\textit{Decreasing commitment}} \\
vLLM & 99.64 & $+5.26$ & 99.77 & $+11.11$ & 99.86 & $+17.65$ \\
\sys & \textbf{99.64} & \textbf{$-0.01$} & \textbf{99.81} & \textbf{$-0.05$} & \textbf{99.82} & \textbf{$+1.08$} \\
DynamoLLM & 76.34 & $-16.13$ & 81.81 & $-12.71$ & 83.22 & $-9.34$ \\
Static-Target & 99.70 & $+4.37$ & 99.82 & $+9.43$ & 99.90 & $+14.93$ \\
Fixed-Config & 99.39 & $+2.84$ & 99.68 & $+8.55$ & 99.78 & $+14.94$ \\
\bottomrule
\end{tabular*}
\end{table}

\begin{figure}[htbp]
 \centering
 \includegraphics[width=\textwidth]{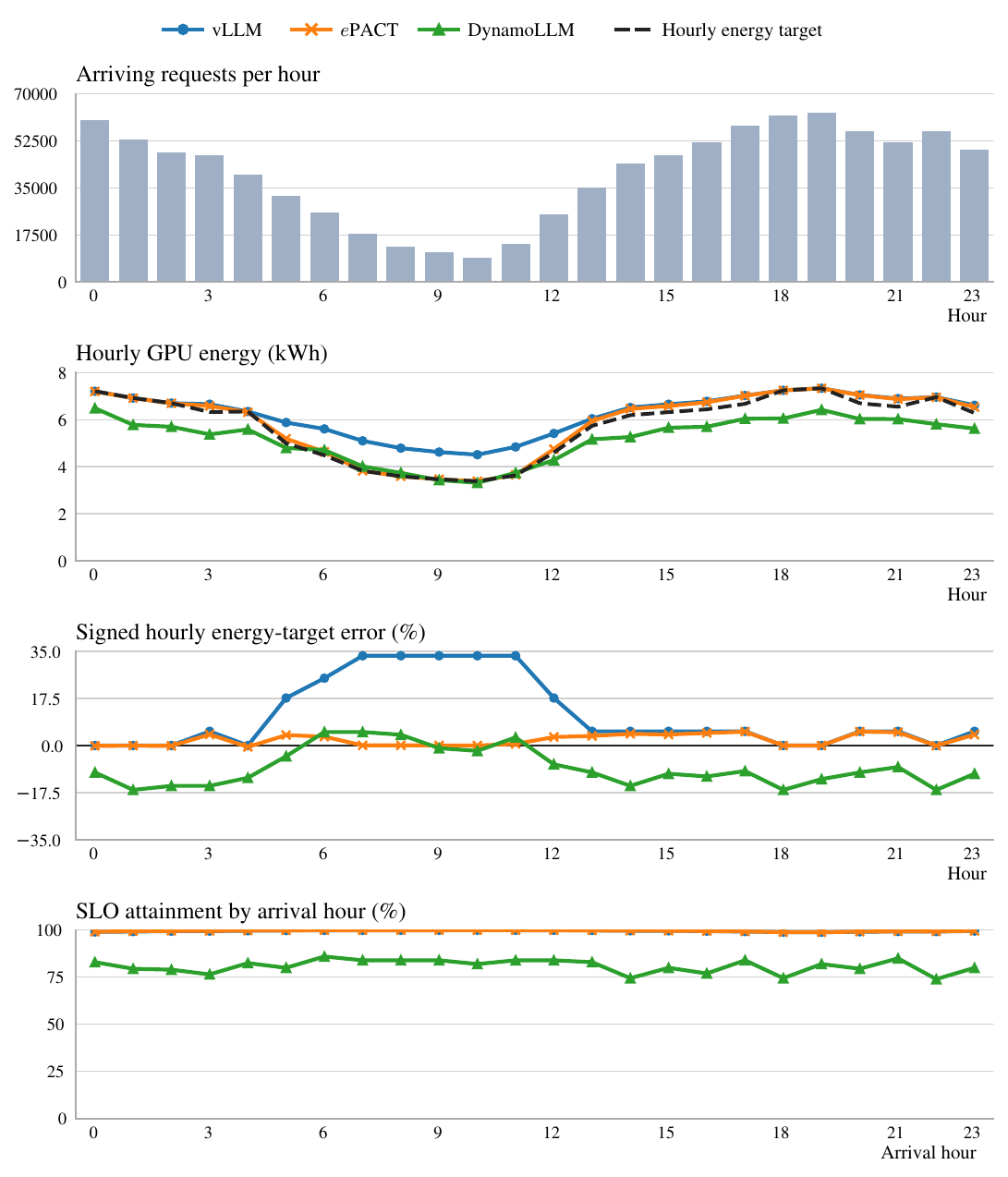}
 \caption{Hourly overview of the 24-hour H20 comparison at fixed $3\times$ SLO. The panels show arrivals, GPU energy, signed commitment deviation, and SLO attainment by arrival hour.}
 \label{fig:h20_day_overview}
\end{figure}

\begin{table}[!ht]
 \centering
 \caption{\sys's hourly H20 results at SLO scale $3$ on the May~11,~2024 trace. Target fractions refer to the historical vLLM calibration. Attainment groups requests by arrival hour; positive deviation indicates overuse. Hourly values are rounded as reported.}
 \label{tab:hourly_tracking}
 \vspace{4pt}
 \begin{tabular*}{\linewidth}{@{\extracolsep{\fill}}rrrrr@{}}
  \toprule
  Hour & Target fraction & Target (MJ) & Attainment (\%) & Deviation (\%) \\
  \midrule
  00 & 100\% & 25.928 & 99.15 & $-0.12$ \\
  01 & 100\% & 24.889 & 99.41 & $-0.04$ \\
  02 & 100\% & 24.111 & 99.60 & $-0.14$ \\
  03 & 95\% & 22.747 & 99.62 & $+4.10$ \\
  04 & 100\% & 22.833 & 99.81 & $-0.52$ \\
  05 & 85\% & 17.975 & 99.92 & $+3.87$ \\
  06 & 80\% & 16.133 & 99.96 & $+3.29$ \\
  07 & 75\% & 13.758 & 99.97 & $+0.05$ \\
  08 & 75\% & 12.923 & 99.99 & $+0.06$ \\
  09 & 75\% & 12.471 & 100.00 & $+0.03$ \\
  10 & 75\% & 12.175 & 100.00 & $-0.01$ \\
  11 & 75\% & 13.066 & 99.99 & $+0.58$ \\
  12 & 85\% & 16.552 & 99.97 & $+3.14$ \\
  13 & 95\% & 20.643 & 99.87 & $+3.55$ \\
  14 & 95\% & 22.272 & 99.72 & $+4.33$ \\
  15 & 95\% & 22.737 & 99.72 & $+4.08$ \\
  16 & 95\% & 23.164 & 99.52 & $+4.66$ \\
  17 & 95\% & 24.009 & 99.29 & $+5.12$ \\
  18 & 100\% & 26.066 & 98.95 & $-0.01$ \\
  19 & 100\% & 26.382 & 98.90 & $-0.04$ \\
  20 & 95\% & 24.095 & 99.20 & $+5.14$ \\
  21 & 95\% & 23.560 & 99.37 & $+4.93$ \\
  22 & 100\% & 25.027 & 99.31 & $-0.08$ \\
  23 & 95\% & 22.592 & 99.66 & $+4.03$ \\
  \bottomrule
 \end{tabular*}
\end{table}

\begin{figure}[htbp]
 \centering
 \includegraphics[width=\textwidth]{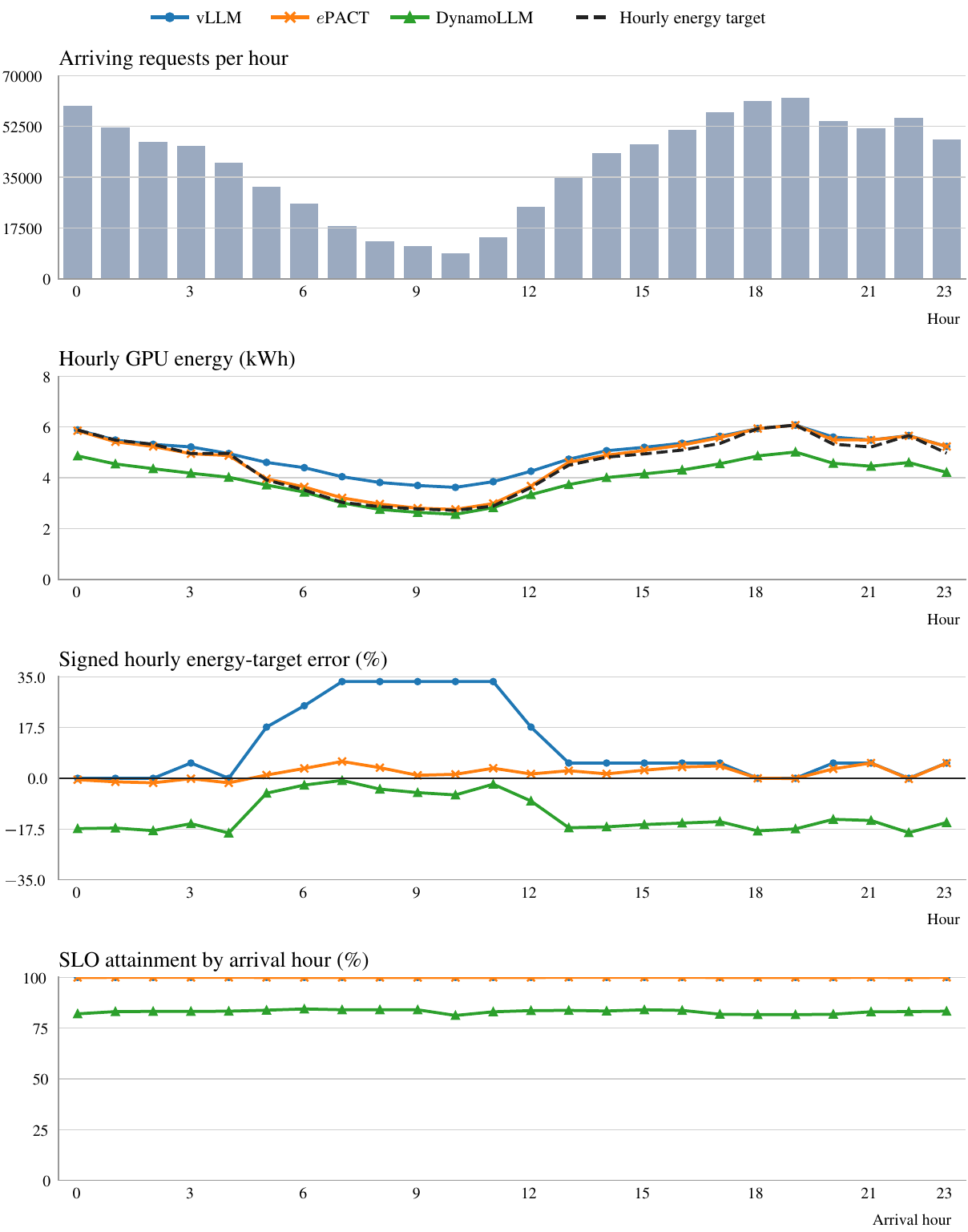}
 \caption{Hourly overview of the 24-hour H200 comparison at fixed $3\times$ SLO. The panels show arrivals, GPU energy, signed commitment deviation, and SLO attainment by arrival hour.}
 \label{fig:h200_day_overview}
\end{figure}

\begin{table}[!ht]
 \centering
 \caption{\sys's hourly H200 results at SLO scale $3$ on the May~11,~2024 trace. Target fractions scale the offline evaluation-day vLLM reference. Attainment groups requests by arrival hour; positive deviation indicates overuse. Hourly values are rounded as reported.}
 \label{tab:h200_hourly_tracking}
 \vspace{4pt}
 \begin{tabular*}{\linewidth}{@{\extracolsep{\fill}}rrrrr@{}}
  \toprule
  Hour & Target fraction & Target (MJ) & Attainment (\%) & Deviation (\%) \\
  \midrule
  00 & 100\% & 21.176 & 99.92 & $-0.49$ \\
  01 & 100\% & 19.742 & 99.95 & $-1.22$ \\
  02 & 100\% & 19.137 & 99.97 & $-1.56$ \\
  03 & 95\% & 17.833 & 99.94 & $-0.13$ \\
  04 & 100\% & 17.833 & 99.99 & $-1.57$ \\
  05 & 85\% & 14.091 & 99.92 & $+1.12$ \\
  06 & 80\% & 12.664 & 99.91 & $+3.39$ \\
  07 & 75\% & 10.911 & 99.93 & $+5.80$ \\
  08 & 75\% & 10.296 & 99.91 & $+3.66$ \\
  09 & 75\% & 9.978 & 99.89 & $+1.03$ \\
  10 & 75\% & 9.777 & 99.86 & $+1.37$ \\
  11 & 75\% & 10.382 & 99.93 & $+3.46$ \\
  12 & 85\% & 13.028 & 99.92 & $+1.49$ \\
  13 & 95\% & 16.184 & 99.98 & $+2.61$ \\
  14 & 95\% & 17.324 & 99.97 & $+1.53$ \\
  15 & 95\% & 17.785 & 99.97 & $+2.79$ \\
  16 & 95\% & 18.328 & 99.98 & $+3.88$ \\
  17 & 95\% & 19.265 & 99.92 & $+4.26$ \\
  18 & 100\% & 21.374 & 99.89 & $-0.00$ \\
  19 & 100\% & 21.867 & 99.89 & $-0.07$ \\
  20 & 95\% & 19.162 & 99.92 & $+3.25$ \\
  21 & 95\% & 18.764 & 99.97 & $+5.26$ \\
  22 & 100\% & 20.379 & 99.91 & $-0.20$ \\
  23 & 95\% & 17.896 & 99.97 & $+5.26$ \\
  \bottomrule
 \end{tabular*}
\end{table}

\FloatBarrier
\subsection{Prediction Accuracy and Control Overhead}
\label{sec:local_eval}
\label{app:eval_validation}

We validate frozen prediction models on held-out workloads and report planning time.

\noindent\textbf{Validation without refitting.}
The first validation uses five workload classes excluded from fitting, repeated twice, on a single replica at 1980\,MHz with fixed batches. Class-level end-to-end latency WAPE ranges from $0.44\%$ to $2.66\%$, and energy WAPE ranges from $0.19\%$ to $2.89\%$. The largest request-level relative latency error among 18 requests is $3.12\%$; the worst evaluated energy cell has $3.28\%$ error. The duration and power models are frozen before evaluation and are not refitted using the held-out measurements.

\noindent\textbf{Continuous-arrival validation.}
A second validation uses 97 requests from a nonoverlapping slice of the same trace, excluded from the corresponding fit, with continuous arrivals and continuous batching on one replica. End-to-end latency WAPE is $4.15\%$, the 95th percentile of per-request relative prediction error is $14.74\%$, and 84 of 97 predictions fall within $\pm5\%$ of measured latency. The signed energy error over the 120-second window is $+0.76\%$. These checks characterize the stated single-replica settings; they do not establish an error bound for every 32-replica configuration or full-day trajectory.

\noindent\textbf{Interpreting prediction errors.}
For measured values $y_j$ and predictions $\widehat{y}_j$, WAPE is $100\frac{\sum_j|\widehat{y}_j-y_j|}{\sum_j y_j}$, computed separately for latency and energy. It is an aggregate error, not the fraction of requests within $5\%$ error. The continuous-arrival energy result is a signed window error, not an energy-WAPE measurement across multiple windows.

\noindent\textbf{Observed control and batching behavior.}
The reported \sys run completes 60 asynchronous decisions with a mean search wall time of $23.35$ seconds. Serving continues while the search runs; this measurement does not by itself establish that planning has no latency impact. Batch sizes 1, 2, 3, and 4 occur in 265{,}295, 113{,}992, 38{,}503, and 1{,}199 steps, respectively, confirming that execution is not restricted to batch size one.

\fi

\end{document}